\documentclass[9pt,twocolumn,twoside]{osajnl}
\journal{ol} % Choose journal (ao, aop, josaa, josab, ol, pr)

\setboolean{shortarticle}{true}
\title{A Super-resolution Optical Classifier with High Photon Efficiency}% using Photon Statistics
\author[1,2]{He Zhang}
\author[1,2]{Santosh Kumar}
\author[1,2,*]{Yu-Ping Huang}

\affil[1]{Department of Physics, Stevens Institute of Technology, Hoboken, NJ, 07030, USA}
\affil[2]{Center for Quantum Science and Engineering, Stevens Institute of Technology, Hoboken, NJ, 07030, USA}

\affil[*]{Corresponding author: yuping.huang@stevens.edu}

 \date{\today}
 
\begin{abstract}
We propose and demonstrate a photon-efficient optical classifier to overcome the Rayleigh limit in spatial resolution. It utilizes mode-selective sum frequency generation and single-pixel photon detection to resolve closely spaced incoherent sources based on photon counting statistics. Super-resolving and photon efficient, this technique can find applications in microscopy, light detection and ranging (LiDAR), and astrophysics.
\end{abstract}
\begin{document}
\maketitle
%\section{Introduction}
Resolving closely-spaced point sources below the diffraction limit has always been an active pursuit in many branches of science and technology, especially those of biomedical imaging \cite{hell_breaking_1994,hell_far-field_2007} and astrophysics \cite{zmuidzinas_cramerrao_2003,Serabyn_2010}. For a long time, the ultimate imaging resolution achievable by coherent optics had been believed to abide the Rayleigh criterion \cite{Rayleigh}, albeit that it is not rooted in the quantum mechanics \cite{Helstrom_1976,helstrom_resolution_1973}. For example, it was shown that the direct measurement of the photon statistics can boost the spatial resolution passing the Rayleigh limit \cite{ober_localization_2004}, as allowed by the Fisher information and photon shot-noise \cite{ram_beyond_2006,wilt_photon_2013}. This can be done by simple, direct measurement of light intensity, although it discards the rich information carried by the photons' phase profiles. Recently, an indirect detection technique was proposed based on spatial-mode demultiplexing  \cite{tsang_PRX_2016,tsang_quantum_2019}, where such phase information has shown to provide superior resolution compared to direct measurement \cite{paur_achieving_2016,tham_beating_2017,zhou_quantum-limited_2019,len_resolution_2020}. 

While those exciting developments were made in the linear optics domain, nonlinear optical techniques have grasped notable progress in applications such as spiral phase-contrast imaging \cite{qiu_spiral_2018}, field-of-view enhancement \cite{liu_up-conversion_2019}, lossless photon shaping \cite{koprulu_lossless_2011}, and LiDAR applications \cite{QPMS2017,Rehain2020}, promising advantages over their linear optical counterparts. In particular, sum-frequency (SF) generation has recently shown potential in beating the Rayleigh's curse in time-frequency domain to estimate sub-pulse-width separations \cite{donohue_quantum-limited_2018}, with prospective applications in optical vortex coronagraphy \cite{n-OVC} and resolving small angular separation between a star and a planet \cite{tamburini_overcoming_2006,mari_sub-rayleigh_2012}. 

In this letter, we propose and demonstrate a nonlinear optical technique, based on mode-selective SF generation \cite{QPMS2017,Rehain2020,Santosh19,zhang_mode_2019}, for super-resolution optical classification using photon counting. By using a spatial-light modulator (SLM) to create optimized pump for SF generation, light sources spaced well below the Rayleigh criterion can be faithfully resolved. This nonlinear optical classification technique could find values in LiDAR \cite{li_super-resolution_2020} and mid-infrared upconversion detection \cite{junaid_video-rate_2019}, and lead to quantum sensing and imaging \cite{pirandola_advances_2018,tenne_super-resolution_2019} in biological and interstellar systems \cite{Babu_2012}.

%\section{Model}

\begin{figure}[htbp]
    \centering
    \includegraphics[width=\linewidth]{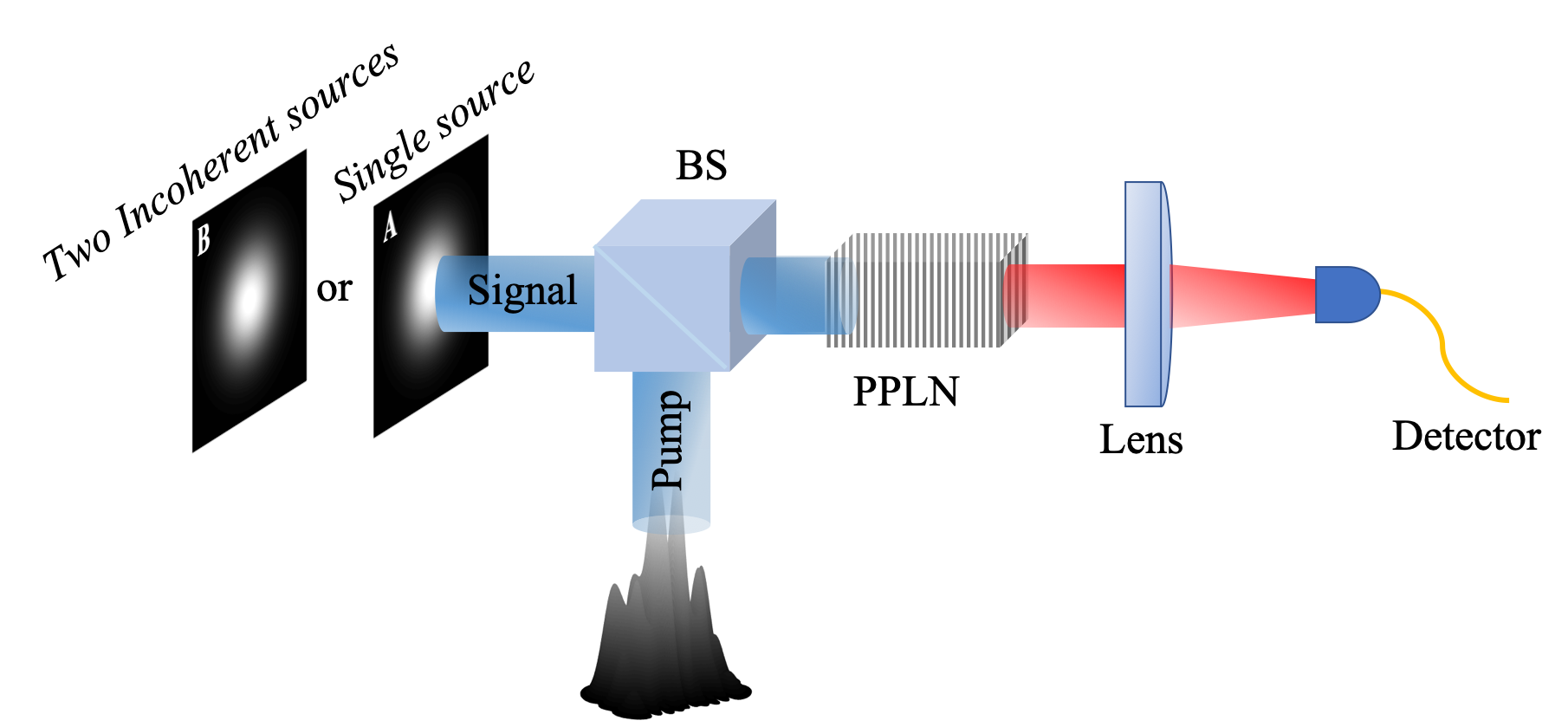}
    \caption{Conceptional illustration of the super-resolution optical classification technique. The target signal, either a single source (case $A$) or two nearby incoherent sources (case $B$), is combined with an optimized pump and focused into a periodic-poled lithium niobate (PPLN) crystal for SF generation. The SF photons are coupled into single mode optical fiber for detection by a single photon counting module.} %the upconverted SF photon number $N^{A}$ and $N^{B}$ (from $A$ and $B$, respectively) to discriminate the incoherent sources. }
    \label{fig:concep-diag}
\end{figure}
A conceptual diagram of the photon-efficient optical classifier is shown in Fig.~\ref{fig:concep-diag}. We aim at distinguishing signals of two kind with the same total illumination: case $A$ from a single source and case $B$ from two equally displaced, incoherent sources with the same brightness. The $A$ signal is in state $\rho_A=|\psi(x,y)\rangle\langle\psi(x,y)|$, while the $B$ signal is in state $\rho_B=\frac{1}{2}\left(|\psi(x+\theta_x,y)\rangle\langle\psi(x+\theta_x,y)|+|\psi(x-\theta_x,y)\rangle\langle\psi(x-\theta_x,y)|\right)$, i.e., consisting of two incoherent sources symmetrically displaced by $\theta_x$ from the centroid. The signal in either case is combined with the pump light and passed through a nonlinear crystal for SF generation and the generated SF photons are counted. 

To distinguish those two cases, the pump is prepared in two spatial modes. One is identical to $\psi(x,y)$, and the other is optimized to selectively up-convert the signal in $\rho_B$ over $\rho_A$, i.e., with a higher SF efficiency for $\rho_B$ than $\rho_A$. By sequentially applying the pump in those two modes, the resulting SF photons are counted and compared for the which-case analysis. 

%Note that these signals are incoherent as the pump interacts separately in time to generate the SF photons. 

% To proceed further, we assume the centroid is known, and the signal $A$ is locates at centroid and the two incoherent sources in case $B$ are symmetric with respect to the centroid.
%The light sources in case $A$ and $B$ are presented use wave function $\psi_{A}(x)$ and $\psi_{B}(x \pm \theta_x)$, respectively.
For a simplified analysis, we assume a Gaussian point spread function (PSF) for the signal, where
\begin{equation}
    \psi(x,y)=\left(\frac{1}{2\pi \sigma_s^2}\right)^{1/2} \exp\left(-\frac{x^2+y^2}{4\sigma_s^2}\right),
    \label{eqn1}
\end{equation}
with $\sigma_s$ the beam width. Accordingly, the two pump modes are each Gaussian in Eq.~(\ref{eqn1})
, and in an optimized superposition of the Hermite-Gaussian (HG) modes as
\begin{equation}
%\psi^P(x)= c_0 \Phi_0(x)+\sum_0^{2l+1} c_{2l+1} \Phi_{2l+1}(x),
\psi^p(x,y)=\sum_{l}\sum_{m} c_{lm} \Phi^p_{lm}(x,y),
\label{pump_mode}
\end{equation}
where 
\begin{equation}
 %\Phi_{l}(x)=\left(\frac{1}{2\pi \sigma^2}\right)^{1/4}\frac{1}{\sqrt{2^l l!}} H_{l}\left(\frac{x}{\sqrt{2}\sigma}\right)\exp\left({-\frac{x^2}{4\sigma^2}}\right),
 \begin{aligned}
    \Phi^p_{lm}(x,y)&=\sqrt{\frac{1}{2\pi \sigma_p^2}}\frac{1}{\sqrt{2^l l!}}\frac{1}{\sqrt{2^m m!}} H_{l}\left(\frac{x}{\sqrt{2}\sigma_p}\right)H_{m}\left(\frac{y}{\sqrt{2}\sigma_p}\right)\\
     &\times\exp\left({-\frac{x^2+y^2}{4\sigma_p^2}}\right).
\end{aligned}
\end{equation}
Here, $\{c_{lm}\}$ are complex coefficients, $\sigma_p$ is the beam width of the pump and $H_j$ is the $j^{th}$-order Hermite polynomial \cite{tsang_PRX_2016}. To achieve high selectivity, we apply an adaptive feedback control loop in our experiment to arrive at the optimal pump profile; see our previous work \cite{zhang_mode_2019}. To minimize the spatial overlap between the optimized pump and signal in $\rho_A$---thus leading to high selectivity---the pump consists of only HG modes with odd $l$'s. As a result, the optimized pump modes are always anti-symmetric of the $x$ axis. They will convert $\pm \theta_x$ displaced Gaussian modes with the same efficiency. As such, in this experiment, we effectively create the signal in $\rho_B$ by using a blaze-grating phase mask on the SLM to displace the Gaussian mode along one direction only. 

%In our experiment, we adaptively prepare the pump in the optimized mode as a superposition of HG modes; see Eq.~(\ref{pump_mode}) \cite{zhang_mode_2019}. 

%In the optical classification, optimized pump retrieved more SF photons from case B, in which the two incoherent sources are separated with respect to the centroid, and only HG modes with odd order $H_l$ has same property of symmetry are excited significantly in that case, we can focus on the superposition of HG modes with odd order $H_l$ to make the optimization more efficient. Here, we choose 20 HG modes as superposed mode basis with $l=1,3,5,7$ and $m=0,1,2,3,4$.

%%and $l =0$ to $2l+1$. In our optimization process, we consider odd higher order modes upto 11.
%%Since the two incoherent sources in case $B$ are symmetry with respect to the center, and the odd order of Hermite-Gaussian modes that has the some property could significantly contribute more in optimization, we can focus on the odd order of Hermite Gaussian modes to prepare the optimized pump. 
%Therefore, the optimized pump mode should has same symmetric character, which brings out the same SF result either of the two incoherent sources in case B. 

%In this paper, we  with signal $A$ and $B$, respectively %as the SF photon number of signal $B$ to $A$ with
To illustrate our approach, let $G_{A(B)}$ and $O_{A(B)}$ be the number of SF photons created by the pump in the Gaussian and optimized modes, respectively, for the $A$ ($B$) signal. The selectivity by the optimized pump is then defined as $S=O_A/O_B$. By differentiating those numbers, there could be multiple analysis methods. For simplicity, we consider a single-parameter tester as the extinction ratio of the photon counts in those two cases, defined as $R_{i}=O_i/G_i$ with $i\equiv\{A,B\}$.  The pump is optimized to selectively upconvert the signal in $\rho_B$, so that $R_A < R_B$. Assuming shot noise for all photon counts, the condition for a faithful discrimination is \begin{equation}
    R_{A}+\Delta R_{A} < R_{B}-\Delta R_{B},
    \label{eq:cutoff}
\end{equation}
with $\Delta R_{i}=R_i\sqrt{(O_i+G_i)/O_i G_i}$. Then, a threshold ratio $R_t$ can be defined as
\begin{equation}
 R_{t}= (R_A+R_{B}+\Delta R_A-\Delta R_{B})/2.
 \label{eq:Rt}
\end{equation}
If the measured ratio is less than $R_t$, the signal is classified to be $A$. Otherwise it is $B$. 

In this work, we focus on maximizing the photon efficiency, aiming at detecting the minimum number of total photons $N_\textit{min}$ while still able to reliably distinguish the two signals.
To this end, we observe that in our application scenario of interest, the displacement in case $B$ is well below the Rayleigh criteria, so that $G_A\approx G_B$. Furthermore, since the optimized pump selectively converts $\rho_B$, one always has $O_B>O_A$. Hence, for a given selectivity, $N_\textit{min}$ shall be $G_B+O_B$ that minimally satisfy the condition in Eq.~(\ref{eq:cutoff}). When $S\ll 1$, it is obtained when $G_B\approx O_B$, as
\begin{equation}
    N_\textit{min}= \frac{2(\sqrt{2}+\sqrt{R_A(1+R_A)})^2}{(R_A-1)^2},
    \label{eq:N-total}
\end{equation}
Note that this result has only considered the shot noise of the signal photon counting. With the detector dark count, background noises, etc., more photons will need to be detected for a reliable classification.   

\begin{figure*}[!htb]
\centering
\includegraphics[width=11cm]{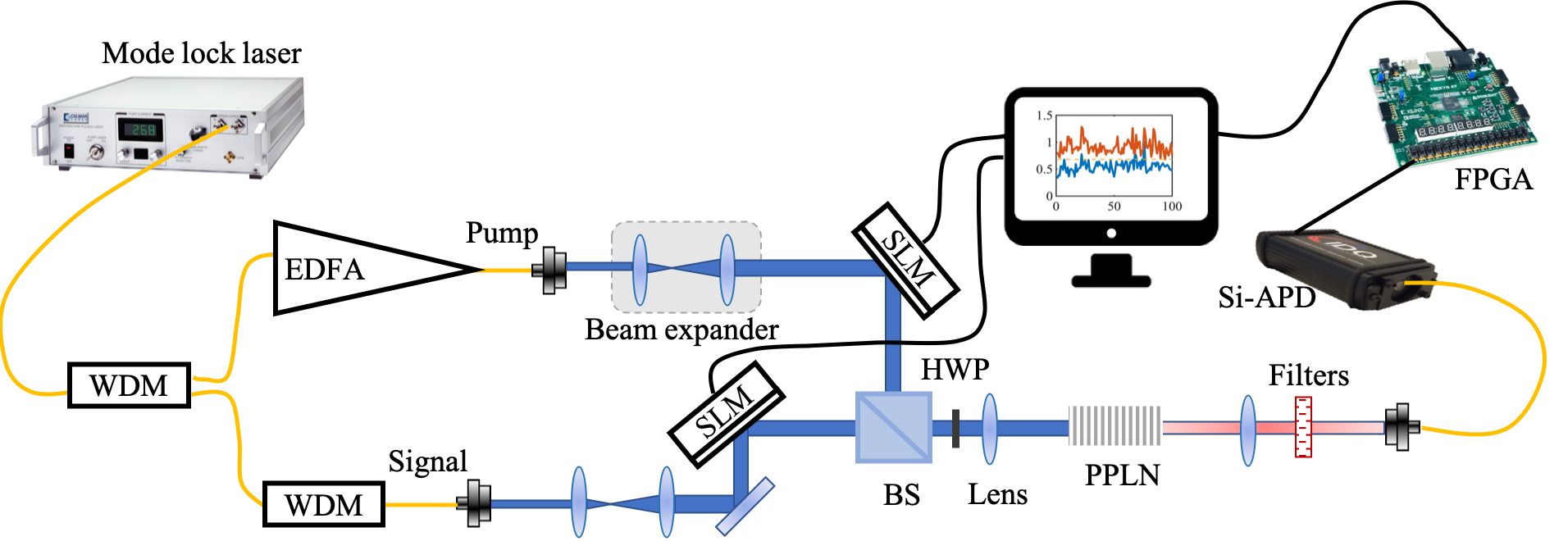}
\caption{Experimental setup. Two pulse trains, each at 1545 and 1558 nm, are generated from a mode lock laser and filtered by two WDM filters. The SLMs are used to create the phase masks for generating shifted incoherent optical signals and optimum pump modes. The pulse trains are combined using a beamsplitter and passed through a temperature stabilized PPLN crystal. The SF light at 775.5 nm is filtered and coupled into a single mode fiber for photon counting. EDFA: Erbium-Doped Fiber Amplifier, QWP: Quarter Waveplate, HWP: Half Waveplate, BS: Beamsplitter, SLM: Spatial Light Modulator, PPLN crystal: Magnesium-doped Periodic Poled Lithium Niobate crystal, Si-APD: Silicon Avalanche Photodiode, TC: Temperature Controller. }
\label{fig:Exp}
\end{figure*}

Figure \ref{fig:Exp} outlines our experimental setup. A mode-lock laser is used to generate the pulse train with 6 ps full-width at half maximum (FWHM) and 50 MHz repetition rate. We use two inline narrow-band wavelength division multiplexers (WDM) of 0.8-nm linewidth to pick two separate wavelengths, one at 1545 nm as the pump and another at 1558 nm as the signal. The pump optical pulses are then amplified in an Erbium-doped fiber amplifier to obtain 6 mW average power with a pulse energy $\sim$ 0.12 nanojoules. The signal optical pulses, on the other hand, are attenuated using neutral density filters to emulate ultrafaint light intensity received. We use free space optics on both arms to select the horizontal polarization for the pump and signal beams. The transverse FWHM of the pump and signal beams are 2.8 mm and 2.6 mm, respectively. Then, these two beams are incident on separate spatial-light modulators (Santec SLM-100, 1440 $\times$ 1050 pixels, pixel pitch 10.4 $\times$ 10.4 $\mu$m) at $50^{\circ}$ and $55^{\circ}$ incidence angle, respectively \cite{Santosh19}. 
A grating phase mask is uploaded onto the signal SLM to displace the input Gaussian beam from the centroid. The pump SLM is used to manipulate and find the optimum phase mask for selectively upconverting the signals. In this experiment, the pump is prepared an optimized superposition of 20 HG modes with $l=1,3,5,7$ and $m=0,1,2,3,4$. Including more HG modes will improve the selectivity and thus the photon efficiency. 

After their combination upon a beamsplitter, a lens focuses the two beams (focus length $F = $200 mm) into a temperature-stabilized second-order nonlinear crystal with a poling period of 19.36 $\mu$m (5 mol.\% MgO doped PPLN, 10 mm length, 3 mm width, and 1 mm height) for SF generation. A half wave plate, before the nonlinear crystal, is used to ensure the vertically polarized light parallel to the crystal's optical axis where the SF generation process is quasi phase matched. The beam width of pump ($\sigma_p$) and signal ($\sigma_s$), inside the crystal, are 22.5 $\mu$m and 20.5 $\mu$m, respectively. The output pulses are then filtered with three short-pass filters which provide a total $>$180 dB extinction to remove any residual fundamental lights \cite{QPMS2017}. The output SF light is coupled into a single-mode fiber (SMF-28) using a fiber collimator consisting of aspheric lens (Thorlabs A375TM-B) and then detected by a silicon avalanche photodiode (Si-APD, ID100-SMF20). The dark-count rate of the detector is about 3 Hz. The photon statistics is then retrieved by the single photon counting performed by the field programmable gate array (FPGA, Zedboard ZYNQ-7000) and sent to a computer for post processing. % through a MATLAB interface.

\begin{figure*}[!htb]
    \centering
    \includegraphics[width=12cm]{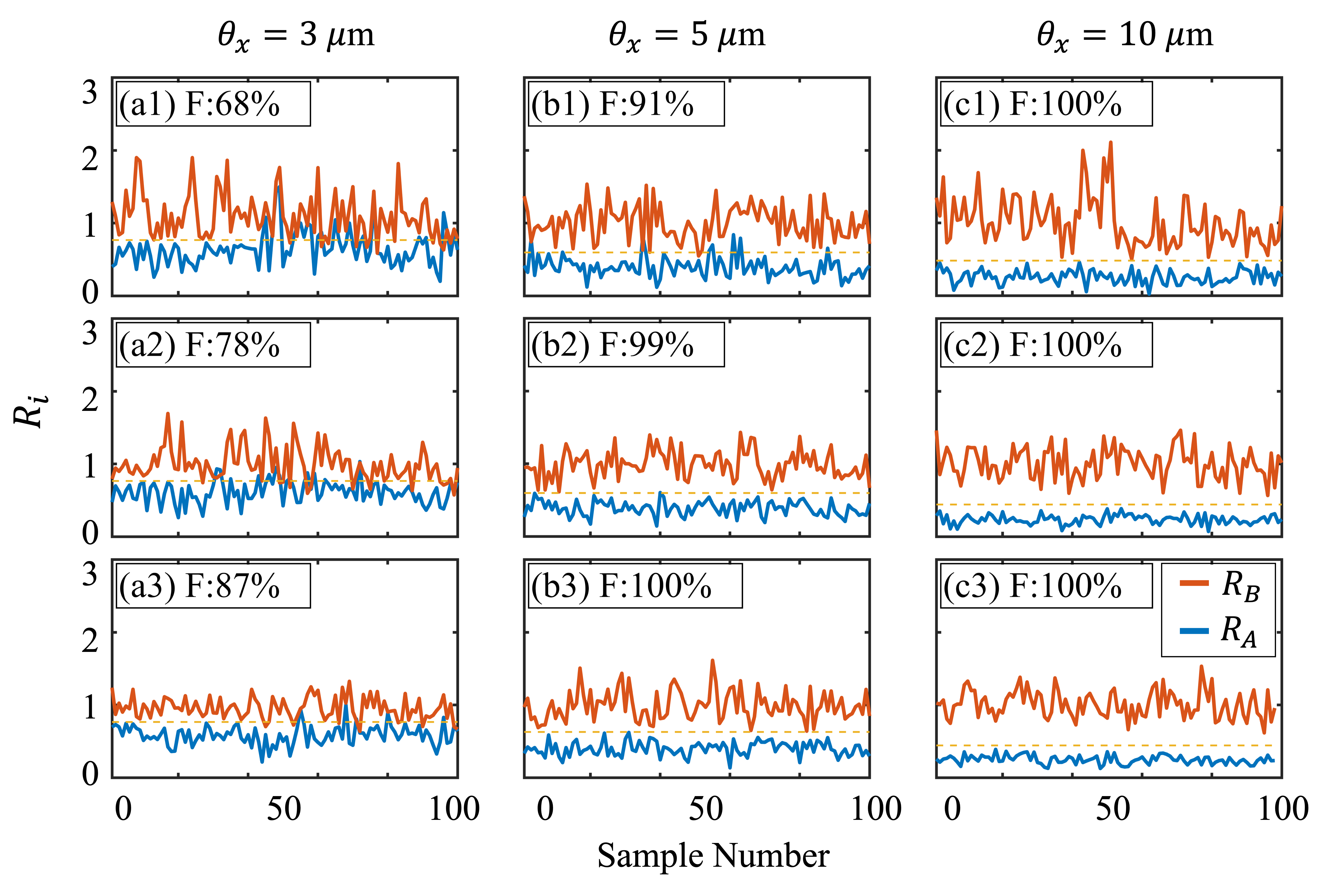}
    \caption{Experimental results for $\theta_x$ = 3 $\mu$m (a1-a3), 5 $\mu$m (b1-b3), and 10 $\mu$m (c1-c3). The first, second, and third row are for $N_\textit{ave}$ around 60, 110, and 170, respectively. In each subplot, F is the test fidelity, and $R_A$, $R_B$, and $R_t$ are drawn in red, blue, and yellow, respectively.}
    \label{fig:hypo-test}
\end{figure*}

%with the conversion extinction ratio $\alpha =$ 1.6, 2.5 and 3.4
%In addition, the extinction ratio of the Gaussian pump are constant for these three separations, where $\beta\approx 1$.
% In our experiment, we choose to roughly match $O_B$ and $G_B$, so as to minimize the detected photon number needed to satisfy Eq.~(\ref{eq:cutoff}). 
In our experiment, we test three displaced signals: $\theta_x$=$3 \mu$m, $5 \mu$m, and $10 \mu$m. In each case, the minimum photon count $N_\textit{min}$ is estimated by Eq.~(\ref{eq:N-total}) to be 95, 36, and 22, with the threshold ratio $R_t$ calculated using Eq.\ref{eq:Rt} as 0.77, 0.61, and 0.45. For those parameters, the classification fidelity, defined as the probability of a correct classification, is measured to be 75\%, 79\%, and 87\%, respectively, which comports with our theory. To increase the fidelity, more photons need to be detected to suppress the shot noise. Figure \ref{fig:hypo-test} plots the results with varied photon counts, where the averaged counts $N_\textit{ave}=\langle G_B +O_B \rangle$ are around 60,~110,~170 in the top to bottom rows, respectively, for each displacement. As seen, a larger photon count suppresses the statistical fluctuation in $R_A$ and $R_B$, making them more distinguishable, leading to a higher fidelity. For the same photon counts, when $\theta_x$ gets smaller, the fidelity drops, but can recovered by higher photon counts. For example, in Fig.~\ref{fig:hypo-test} (a1-a3), as the average detected photons increase from 60 to 170, the classification fidelity improves from 68\% to 87\%. A similar behavior is observed for $\theta_x$=5 $\mu$m, as shown in \ref{fig:hypo-test} (b1-b3), whose fidelity increases from 91\% to 100\%. When $\theta_x$=10 $\mu$m, the fidelity quickly saturates at 100\%, while $R_A$ and $R_B$ curves become further separated as $N_\textit{ave}$ increases.

\begin{figure}[htbp]
     \centering
     \includegraphics[width=7cm]{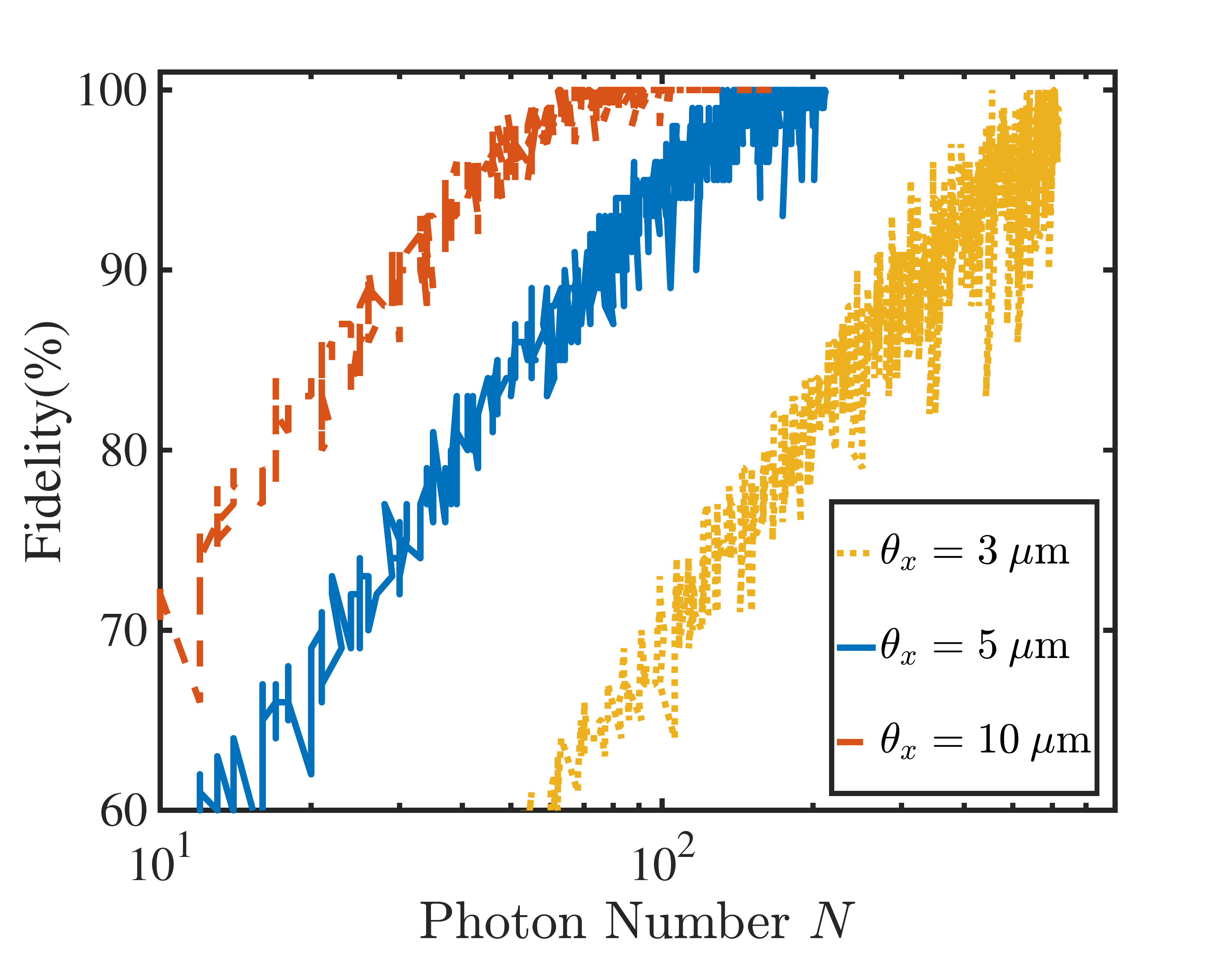}
     \caption{Fidelity versus photon counts with different separations between two incoherent sources. The orange, blue and yellow color represents the results for $\theta_x=$10 $\mu$m, 5 $\mu$m and 3 $\mu$m, respectively.}
     %The solid points are the experimental result with error bar and the solid line are the curve fittings. 
     \label{fig: phoN_fidelity}
 \end{figure}
 
In Fig.~\ref{fig: phoN_fidelity}, we aggregate more experimental results and plot the fidelity versus the average detected photon number $N_\textit{ave}$ for each displacement. In all three cases, a linear increase in the fidelity requires exponential more photon counts. For $\theta_x$=3, 5, and 10 $\mu$m, the fidelity saturates at $N_\textit{ave}=$ 534, 132 and 62, respectively.   
%For classical direct imaging method to estimate two incoherent signals with separation $\theta_x$, the probability density of the position of arrival photons is expressed in terms of the mean intensity as $\Lambda(x)=\frac{1}{2}\left(I(x+\theta_x/2)+I(x-\theta_x/2)\right)$, where $I(x\pm\theta_x/2)\approx I(x)\pm\frac{\theta_x}{2} I'(x)+\frac{\theta_x^2}{8} I''(x)$ and $I(x)=|\psi(x)|^2$ for $\theta_x\ll1$. It gives $\Lambda(x)\approx I(x)+\frac{\theta_x^2}{8} I''(x)$.

\begin{figure}[htbp]
    % \centering
    \includegraphics[width=4.5cm]{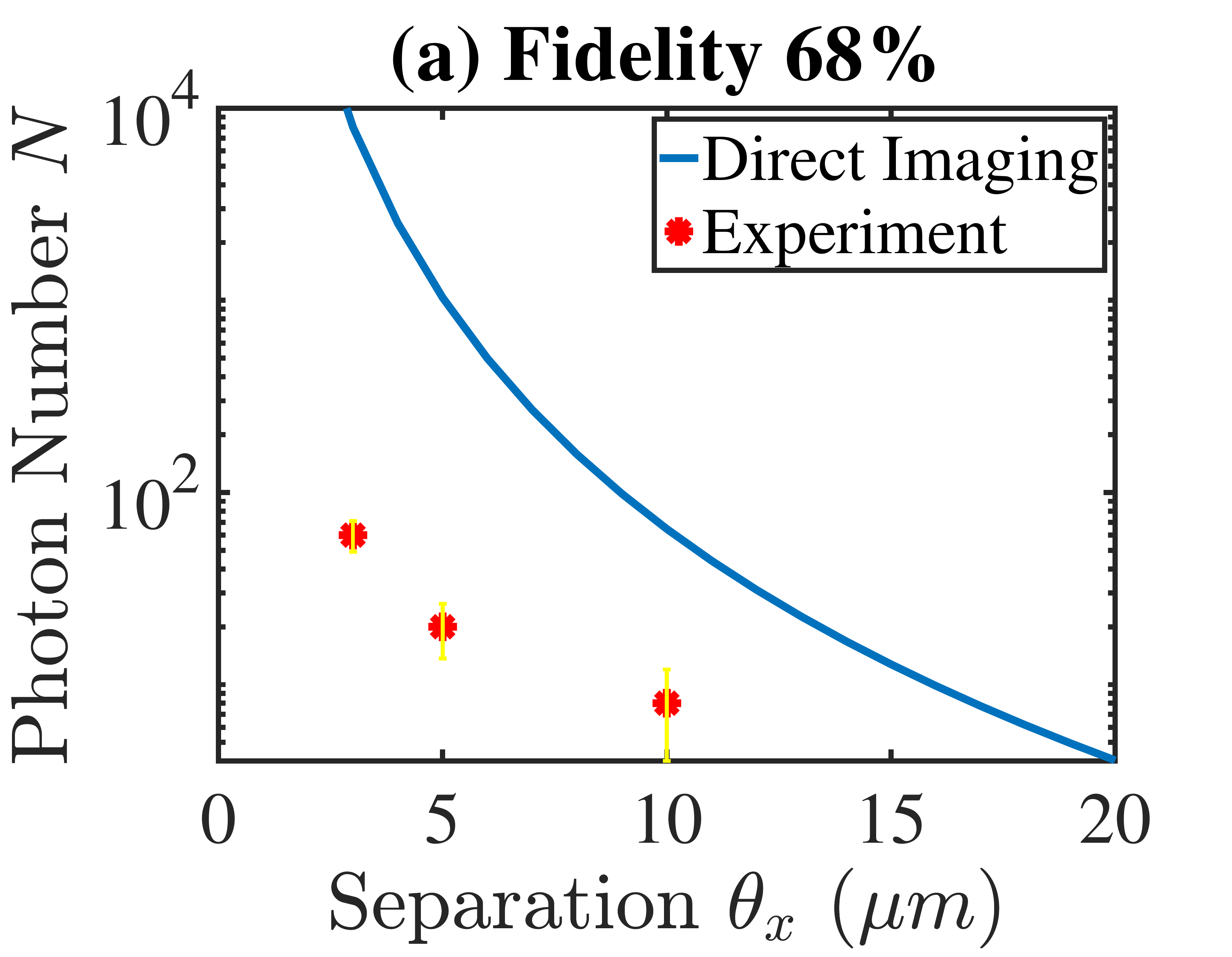}
    \includegraphics[width=4.5cm]{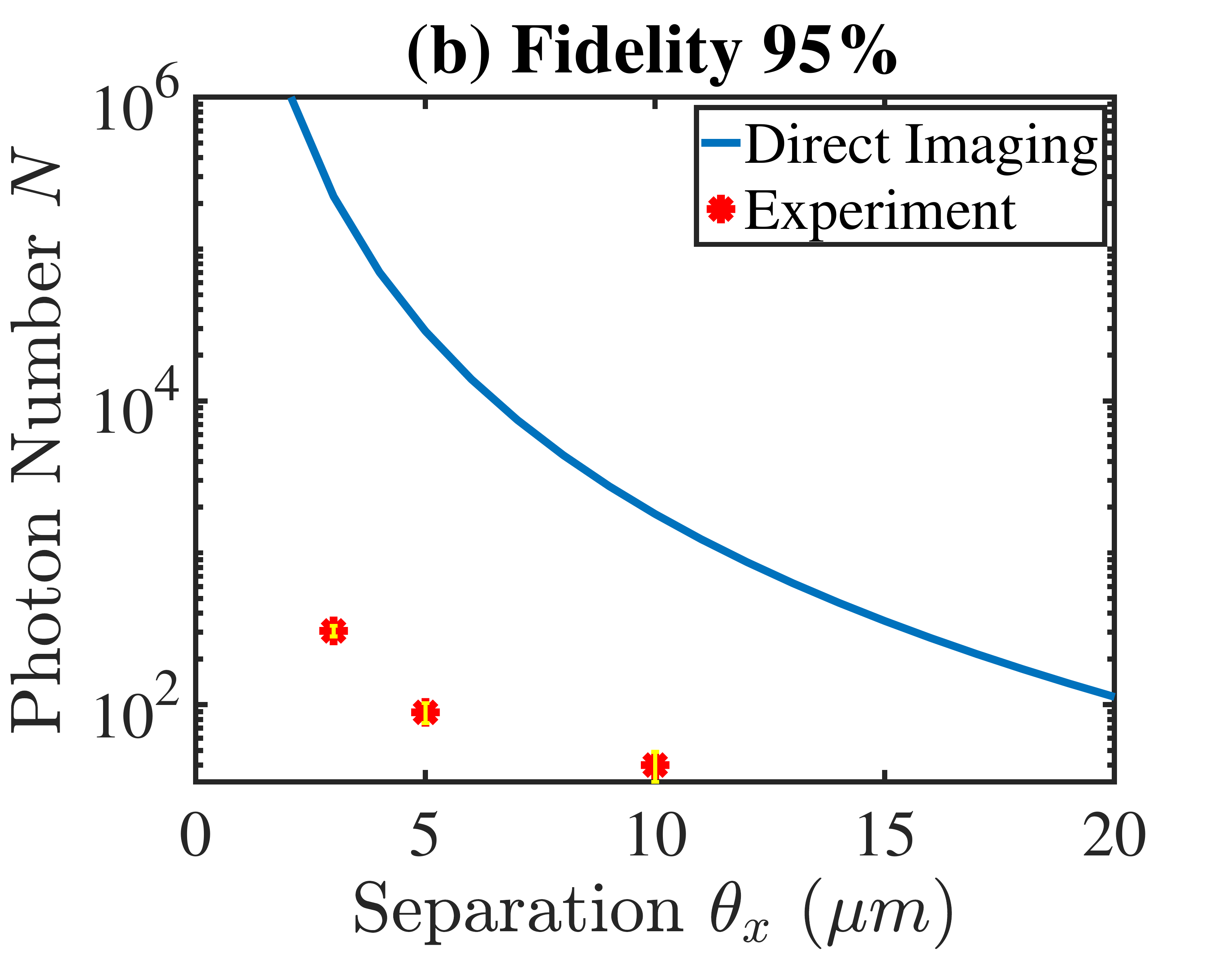}
    \caption{Required photon number to achieve (a) 68\% and (b) 95\% fidelity as a function of $\theta_x$. The blue curves show the theoretical results of ideal direct detection and orange dots mark the present experimental results based on mode selective SF generation. }
    \label{comp-classical}
\end{figure}

Finally, we benchmark the present super-resolution classifier against its direct-detection counterpart. With photon starving applications in mind, we contrast their photon efficiency as a function of the displacement. That's, we compare the required photon counts in each technique to achieve the same fidelity. The results are plotted in Fig. \ref{comp-classical} (a) and (b), showing our experimental results and the theoretical curves of the direct detection method to achieve 68\% and 95\% fidelity, respectively, as a function of $\theta_x$. The direct detection curve is calculated from the Fisher information for direct intensity measurement of two incoherent signals, evaluated under the condition of $\theta_x < \sigma_s \approx$ 20 $\mu$m to be 
\begin{equation}
    \mathcal{F} \approx  \frac{N\theta_x^2}{16}\int_{-\infty}^{\infty} \frac{[I''(x)]^2}{I(x)}dx,
%N\int_{-\infty}^{\infty}\frac{1}{\Lambda(x)}\left(\frac{\partial \Lambda(x)}{\partial \theta_x}\right)^2dx
\end{equation}
with $I(x)=|\psi(x)|^2$ and $N$ the total number of detected photons \cite{paur_achieving_2016,paur_tempering_2018}. %Same as our hypothesis test, we assume that the centroid is known for classical direct imaging.
%and test our hypothesis for discriminating small separated signals. 
For the present Gaussian PSF, the Cram\'er-Rao lower bound for the variance of the estimated $\theta_x$ for $N$ detected photons is given as \cite{paur_tempering_2018},
\begin{equation}
    %\sqrt{Var[\theta_x]}\geq \sqrt{\mathcal{F}^{-1}}=\sqrt{\frac{2}{N}}\frac{2\sigma^2}{\theta_x},
     %=\frac{1}{N}\frac{\theta_x^2 k_2+\theta_y^2 k_1}{(\theta_x^2 -\theta_y^2) k_1 k_2}\\
     Var[\theta_x]\geq \frac{1}{N\mathcal{F}}=\frac{8\sigma^4}{\theta_x^2N}.
     \label{7}
\end{equation}
Assuming a normal probability distribution, a classification fidelity of 68\% is achieved at $Var[\theta_x]=4.8\theta_x^2$, and a 95\% fidelity requires $Var[\theta_x]\approx 0.4\theta_x^2$. Then, $N$ as a function of $\theta_x$ can be evaluated using Eq.~(\ref{7}). These results are shown in Fig. \ref{comp-classical} in blue lines, compared with our experimental results shown in red dots. As seen, the required photon counts for different displacement using our optical classifier is much smaller than the theoretical limit of the direct detection. For a small $\theta_x$, the photon efficiency is improved by 100 to 1000 times. It clearly shows that the mode-selective nonlinear classifier can overcome the Rayleigh’s curse in resolving incoherent, nearby light sources, making it valuable for photon starving applications where the signal photons are rare or the photon detection duration is restrictively short. %*******He/Santosh, we need to give $\sigma$ value somewhere, maybe after the experimental setup **********
%Here, these two separated incoherent signals are symmetry,
%these two separated incoherent signals are symmetry, we consider the fisher information of estimating $\theta_x$ of two symmetry signals  ($\Lambda(x)=\frac{1}{2}(I(x+\theta_x/2)+I(x-\theta_x/2))$) is same as one signal ($\Lambda(x)=I(x+\theta_x)$). In our hypothesis test, instead of sending two symmetry signals on both sides of the cetroid point at same time, we consider one signal at every time. 
%Fig. \ref{comp-classical} shows the comparison of the needed photon number for estimating two signals between direct imaging method and experimental hypothesis test result.

%\section*{Conclusion}
In summary, we have demonstrated a nonlinear optical approach to resolving incoherent sources whose separation is well below the Rayleigh resolution limit, achieving orders of magnitude advantages in photon efficiency. It employs mode selective frequency conversion in the spatial domain to pick up the salient features in the optical modes that are not distinguishable in their intensity profiles. While only the transverse displacement is considered in this work, the same technique can be applied to axial displacement and other super-resolution imaging tasks. It can find also applications in quantum sensing and imaging for astronomy \cite{tamburini_overcoming_2006,mari_sub-rayleigh_2012,Babu_2012} and biology \cite{,tenne_super-resolution_2019}.

%\section*{Funding Information}
%\noindent\textbf{Funding.} This research was supported in part by National Science Foundation (NSF).%Award numbers 1806523 and 1842680

%\section*{Acknowledgments}

\noindent\textbf{Acknowledgments} The authors thank Jeeva Ramanathan and Patrick Rehain for helps on FPGA photon counting.

%\section*{Disclosures}
\noindent\textbf{Disclosures.} The authors declare no conflicts of interest.

%\section*{Supplemental Documents}

%\bigskip \noindent See \href{link}{Supplement 1} for supporting content.

%\section*{References}

% Bibliography
%\nocite{*}
%%%\bibliography{sample}

% Full bibliography will be added automatically on a new page for Optics Letters submissions. This command is ignored for journal article submissions.
% Note that this extra page will not count against page length.
%%%\bibliographyfullrefs{sample}

%Manual citation list
%\begin{thebibliography}{1}
%\bibitem{Zhang:14}
%Y.~Zhang, S.~Qiao, L.~Sun, Q.~W. Shi, W.~Huang, %L.~Li, and Z.~Yang,
 % \enquote{Photoinduced active terahertz metamaterials with nanostructured
  %vanadium dioxide film deposited by sol-gel method,} Opt. Express \textbf{22},
  %11070--11078 (2014).
%\end{thebibliography}

\end{document}